\documentclass[epj]{svjour}

\usepackage{graphicx}
\usepackage{dcolumn}
\usepackage{bm}
\usepackage{psfrag}
\usepackage{epsfig}
\usepackage{amsmath,amssymb}

\newcommand{\dd}{\text{d}}
\newcommand{\dod}[2]{\frac{\dd #1}{\dd #2}}

\newcommand*{\ket}[1]{\mathopen{|}#1\mathclose{\rangle}}

\newcommand{\Tr}{\operatorname{Tr}}
 
\newcommand*{\im}{\mathrm{i}} 
   
\newcommand*{\ee}{\mathrm{e}}

\begin{document}


\title{Fourier's Law confirmed for a class of small quantum systems}

\author{M.~Michel\inst{1} \and M.~Hartmann\inst{2} \and J.~Gemmer\inst{1} \and G.~Mahler\inst{1}}
\institute{Institute of Theoretical Physics I, %
           University of Stuttgart, %
           Pfaffenwaldring 57, %
           D-70550 Stuttgart, %
           Germany \and %
           Institute of Thechnical Physics - DLR Stuttgart,  %
           Pfaffenwaldring 38-40, %
           D-70569 Stuttgart, Germany 
}

\date{Received: date / Revised version: date}

\mail{mathias@theo1.physik.uni-stuttgart.de}

\abstract{
Within the Lindblad formalism we consider an interacting spin chain coupled locally to heat baths. We investigate the dependence of the energy transport on the type of interaction in the system as well as on the overall interaction strength. For a large class of couplings we find a normal heat conduction and confirm Fourier's Law. In a fully quantum mechanical approach linear transport behavior appears to be generic even for small quantum systems.
}

\maketitle

%
%
\section{Introduction}

Heat conduction in condensed matter, especially in insulators, is a long standing problem which has attracted renewed interest recently. 
In the classical understanding, heat conduction in insulators results from phonon scattering. 
This process is an anharmonic effect of the lattice. 
There are two different scattering types, the normal or N-processes and so called U-processes. N-processes are momentum conserving and cannot affect the heat conduction of the solid state. Only the second type of scattering, the U-processes, should give rise to a finite heat conductivity: 
In such processes momentum is conserved only modulo a reciprocal lattice vector \cite{Ziman2001,Peierls1929,Peierls1955,Michel1976}. For very low temperatures these U-processes rapidly die out and the heat conductivity diverges. 
In this case only impurities may limit the heat conductivity.

Recently, a series of articles have addressed heat conduction and Fourier's law in one dimensional systems. 
It has been found that regular heat conduction (non vanishing local gradient of temperature) does not occur naturally in the Hamilton models considered:   
Most of these have been classical one-dimensional systems coupled via different interactions \cite{Lepri1998,Garrido2001,Savin2002}. 
Note that because of the lack of phonon scattering, a harmonically coupled chain cannot have a finite heat conductance \cite{Saito2000}. It is necessary to take an anharmonic potential into consideration.   

Heat conduction is a problem of non-equilibrium thermodynamics in which equilibrium properties hold at most locally.  
The equilibrium thermodynamics of small systems can be given a quantum mechanical foundation \cite{GemmerOtte2001,Gemmer2002,Gemmer2003} and therefore it is tempting to expect non-equilibrium behavior also to follow directly from quantum mechanics. 
Indeed, in one-dimensional quantum systems strong indications have been found for a normal heat conduction \cite{Saito1996,Saito2003}. 
But it is not clear yet which conditions a system has to fulfill in order to exhibit local thermodynamical behavior. 
Especially the dependency of heat conduction on the type of internal couplings within the system have not been investigated in full detail.
There are systems showing a normal heat conductance \cite{Saito1996,Saito2003} and others where the conductivity diverges \cite{Lepri1998} (vanishing temperature gradient).
Unfortunately, even the concept of local thermodynamical quantities such as local temperature or local energy easily becomes ambiguous.
Nevertheless it is inevitable to discuss these questions carefully in the context of heat conduction models, since the results depend crucially on the underlying concepts.

To analyze the conditions of thermodynamical behavior in small systems we study a one-dimensional quantum system coupled to heat reservoirs.
The building blocks (here $N=4-6$ spins) of the system are internally coupled with different types of interactions. 
Traditionally, thermodynamical behavior is expected for large systems ($N\rightarrow\infty$), smaller quantum systems are thought to tend to deviate from thermodynamical behavior.
However we could find normal heat conduction even for very small systems ($N=4$).
We consider a F\"orster coupling alone and together with a non-resonant coupling and we introduce a random coupling, too, to test the generality of thermodynamical behavior.
Additionally, we investigate the transport behavior with respect to different coupling strengths.
For these various cases we find diverse transport behavior and therefore means to specify conditions a system has to fulfill in order to show a normal transport behavior.

%
%
\section{Model}

We consider an open quantum system, a chain of $N$ subsystems with $n$ levels each, with nearest neighbor interactions (cf.\ \cite{Mahler1998,Otte2000}). 
The hamiltonian $\mathcal{H}$ of the system thus reads
\begin{equation}
  \label{eq:1}
  \mathcal{H}=\sum_{\mu=1}^{N}\mathcal{H}^{(\text{loc})}(\mu)+\frac{\lambda}{I}\sum_{\mu=1}^{N-1}\mathcal{H}^{(\text{int})}_{\mu,\mu+1}.
\end{equation}
Here, the terms of the first sum represent equidistant n-level subsystems. 
The second term accounts for pair-inter\-actions between two adjacent subsystems, with an overall coupling strength $\lambda$. For $\lambda$ to characterize the absolute interaction strength it is necessary to normalize the different interaction types by $I$, the mean absolute value of interaction matrix elements, i.\ e.\
\begin{equation}
  \label{eq:16}
  I^2=\frac{1}{n^{2N}}\;\Tr\Big\{\Big(\sum_{\mu=1}^{N-1}\mathcal{H}^{(\text{int})}_{\mu,\mu+1}\Big)^2\Big\}
\end{equation}

For $n=2$ (spins) it is possible to use as a set of basis operators the Pauli spin operators
$\hat{\sigma}_i$ ($i=0,x,y,z$). In terms of these operators the local hamiltonian of a subsystem $\mu$ with an energy spacing $\Delta E=1$ can be written as
\begin{equation}
  \label{eq:2}
  \mathcal{H}^{(\text{loc})}(\mu)=\frac{1}{2}\hat{\sigma}_z(\mu).
\end{equation}
The spins are coupled with alternative types of next neighbor interactions (see \cite{Mahler1998}): a non-resonant diagonal interaction
\begin{align}
  \label{eq:3}
\mathcal{H}^{(\text{NR})}_{\mu,\mu+1}=C_{\text{NR}}\;\hat{\sigma}_z(\mu)\otimes\hat{\sigma}_z(\mu+1),
\end{align}
a resonant energy transfer interaction (F\"orster-Coupling)
\begin{align}
  \label{eq:15}
  \mathcal{H}^{(\text{F})}_{\mu,\mu+1}&=\frac{C_\text{F}}{2}\Big(\hat{\sigma}_x(\mu)\otimes\hat{\sigma}_x(\mu+1)\notag\\
  &\hspace{2cm}+\hat{\sigma}_y(\mu)\otimes\hat{\sigma}_y(\mu+1)\Big),
\end{align}
where $C_{\text{NR}}$ and $C_{\text{F}}$ can be used to adjust the relative strength, or a totally random next neighbor interaction
\begin{equation}
  \label{eq:4}
  \mathcal{H}^{(\text{R})}_{\mu,\mu+1}=\sum_{i=1}^{3}\sum_{j=1}^{3} p_{ij}\;\hat{\sigma}_i(\mu)\otimes\hat{\sigma}_j(\mu+1)
\end{equation}
with normal distributed random numbers $p_{ij}$ (variance 1).
Note that $p_{ij}$ is independent of $\mu$ (no disorder). The random interaction is supposed to model `` typical interactions '' without any bias.

To model the influence of a heat bath, we use the Lindblad formalism: The standard Liouville equation is supplemented by an incoherent damping term (see, e.\ g.\ , \cite{Mahler1998}):
\begin{align}
  \label{eq:5}
  \dod{\hat{\rho}}{t}&=-\im\;[\mathcal{H},\hat{\rho}]+\mathcal{L}^{(\text{B})}\hat{\rho}\notag\\
  &=\mathcal{L}\hat{\rho}.
\end{align}
For the local coupling of spin $\mu$ of the chain to a bath we expand in terms of raising and lowering operators:
\begin{align}
  \label{eq:6}
  \mathcal{L}^{(\text{B})}\hat{\rho}=\frac{W_{1\rightarrow 0}}{2}(2\hat{\sigma}^{-}_{\mu}\hat{\rho}\hat{\sigma}^{+}_{\mu}-\hat{\rho}\hat{\sigma}^{+}_{\mu}\hat{\sigma}^{-}_{\mu}-\hat{\sigma}^{+}_{\mu}\hat{\sigma}^{-}_{\mu}\hat{\rho})\notag\\
+ \frac{W_{0\rightarrow1}}{2}(2\hat{\sigma}^{+}_{\mu}\hat{\rho}\hat{\sigma}^{-}_{\mu}-\hat{\rho}\hat{\sigma}^{-}_{\mu}\hat{\sigma}^{+}_{\mu}-\hat{\sigma}^{-}_{\mu}\hat{\sigma}^{+}_{\mu}\hat{\rho})
\end{align}
where the first term describes a decay from $\ket{1}\rightarrow\ket{0}$ with rate $W_{1\rightarrow 0}$ and the second from $\ket{0}\rightarrow\ket{1}$ with $W_{0\rightarrow 1}<W_{1\rightarrow 0}$.
The properties of the environment (bath) only enter via these two rates. 

If the bath was brought in contact with a single spin, the spin would relax to the stationary state:
\begin{equation}
  \label{eq:7}
  \hat{\rho}_{\text{stat}}=\frac{1}{W_{1\rightarrow 0}+W_{0\rightarrow 1}}
\begin{pmatrix}
W_{1\rightarrow 0} & 0 \\
0 & W_{0\rightarrow 1}
\end{pmatrix}.
\end{equation}
Interpreting this state as the result of thermal equilibrium, where $\widetilde{T}$ is defined via the Boltzmann distribution
\begin{equation}
  \label{eq:8}
  \frac{W_{0\rightarrow 1}}{W_{1\rightarrow 0}}=\ee^{-\frac{\Delta E}{\widetilde{T}}}.
\end{equation}
Operationally, we can associate the temperature $\widetilde{T}$ with the temperature of the bath $\widetilde{T}(\text{B})=\widetilde{T}$ measured in units of $\Delta E$.
With (\ref{eq:8}) and the normalization $W_{1\rightarrow 0}+W_{0\rightarrow 1}=\lambda^{(\text{B})}$ it is possible to rewrite the rates in terms of the coupling strength $\lambda^{(\text{B})}$ and the bath temperature $\widetilde{T}(\text{B})$: 
\begin{equation}
  \label{eq:13}
  W_{0\rightarrow 1}=\frac{1}{1+\ee^{\frac{1}{\widetilde{T}(\text{B})}}} \lambda^{(\text{B})},\;W_{1\rightarrow 0}=\frac{1}{1+\ee^{-\frac{1}{\widetilde{T}(\text{B})}}} \lambda^{(\text{B})}.
\end{equation}

To measure the local temperature of a spin $\mu$ in the chain we use the mean excitation energy
\begin{equation}
  \label{eq:9}
  T(\mu)=\Tr \{\hat{\rho}(\mu)\mathcal{H}^{(\text{loc})}(\mu)\}.
\end{equation}
of the respective spin ($0\leq T(\mu)\leq 0.5$). $\mathcal{H}^{(\text{loc})}(\mu)$ and thus $T(\mu)$ are defined in units of $\Delta E$, $\hat{\rho}(\mu)$ is the corresponding reduced density operator. This $T(\mu)$ is a well-defined local quantity, irrespective of further couplings \cite{Ziman2001:Grueneisen}. 
For consistency reasons we define the temperature of the bath now as the mean energy a single spin coupled to this bath would have
\begin{align}
  \label{eq:14}
  T(\text{B})&=\Tr \{\hat{\rho}(1)\mathcal{H}^{(\text{loc})}(1)\}\notag\\
&=\frac{W_{0\rightarrow 1}}{\lambda^{(\text{B})}}=\frac{1}{1+\ee^{\frac{1}{\widetilde{T}(\text{B})}}}.
\end{align}

Rather than applying iterative numerical methods (e.g.\ Runge-Kutta) to integrate the Liouville equation, we diagonalize the Liouville operator $\mathcal{L}$ (see eq.\ (\ref{eq:5})) of the whole open system. We thus obtain the exact solution of eq.\ (\ref{eq:5}) at all times and without iterative numerical errors. 

%
%
\section{Single bath}

In order to investigate whether full thermodynamical equilibrium may be reached for a chain model coupled only at one border to a heat reservoir as mentioned in the last section, we consider a system of $N=4$ spins, where the first spin ($\mu=1$) is coupled to a thermal bath. 
The temperature of the bath is still defined by the two constants $W_{1\rightarrow0}$ and $W_{0\rightarrow1}$ as mentioned in the last section ($\widetilde{T}(\text{B})=1\Rightarrow T(\text{B})\approx0.27$ cf.\ (\ref{eq:14})). 
We choose the coupling strength of the first spin to the bath to be weak ($\lambda^{(\text{B})}=0.01$).

\begin{figure}
\psfrag{0}{}
\psfrag{1}{\tiny 1}
\psfrag{2}{\tiny 2}
\psfrag{3}{\tiny 3}
\psfrag{4}{\tiny 4}
\psfrag{N}{$\mu$}
\psfrag{T}{$T$}
\psfrag{TB}{$\scriptscriptstyle T(\text{B})$}
\psfrag{0.27}{\tiny =0.27}
\psfrag{Bath}{\tiny bath}
\psfrag{0.55}{}
\psfrag{0.5}{\hspace{1mm}\tiny 0.5}
\psfrag{0.45}{\hspace{1mm}\tiny 0.45}
\psfrag{0.4}{\hspace{1mm}\tiny 0.4}
\psfrag{0.35}{\hspace{1mm}\tiny 0.35}
\psfrag{0.3}{\hspace{1mm}\tiny 0.3}
\psfrag{0.25}{\hspace{1mm}\tiny 0.25}
\psfrag{l=0.3}{\hspace{1mm}$\scriptstyle\lambda=0.3$}
\psfrag{l=0.01}{\hspace{1mm}$\scriptstyle\lambda=0.01$}
\psfrag{l=10}{\hspace{1mm}$\scriptstyle\lambda=10$}
\epsfig{file=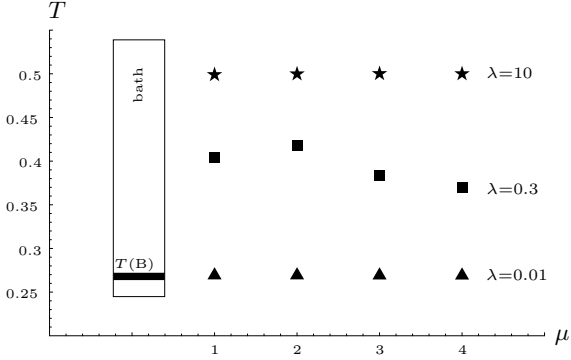,width=8cm}
\caption{Temperature profile: N=4 spins coupled by a random interaction (of internal coupling strength $\lambda=0.01,0.3,10$); first spin $\mu=1$ is coupled to a bath with coupling strength $\lambda^{(\text{B})} = 0.01$ and a temperature $T(B)\approx0.27$ (line in the box marks the temperature of the bath).}
\label{figure1}
\end{figure}

Firstly the coupling between nearest neighbor spins is taken as the random interaction $\mathcal{H}^{\text{(R)}}$. 
From the stationary state of the system we compute the reduced density operators of each spin $\hat{\rho}(\mu)$ and the appropriate temperatures based on eq.\ (\ref{eq:9}). 
The final state is independent of the initial state but not of the internal coupling strength $\lambda$. For very small $\lambda=0.01$ the temperature profile is flat (see FIG.\ \ref{figure1}) while for $\lambda=0.3$ the spins do no longer have the same temperature. 
In case of very strong interactions, $\lambda=10$, the individual spins are in a totally mixed state and the whole system is highly entangled.

\begin{figure}
\psfrag{0.1}{\tiny 0.1}
\psfrag{0.08}{\tiny 0.08}
\psfrag{0.06}{\tiny 0.06}
\psfrag{0.04}{\tiny 0.04}
\psfrag{0.02}{\tiny 0.02}
\psfrag{0.2}{\tiny 0.2}
\psfrag{0.8}{\tiny 0.8}
\psfrag{0.6}{\tiny 0.6}
\psfrag{0.4}{\tiny 0.4}
\psfrag{1}{\tiny 1}
\psfrag{D}{$D_\text{f}$}
\psfrag{D1}{$\scriptstyle D_{\hat{1}}$}
\psfrag{l}{$\lambda$}
\epsfig{file=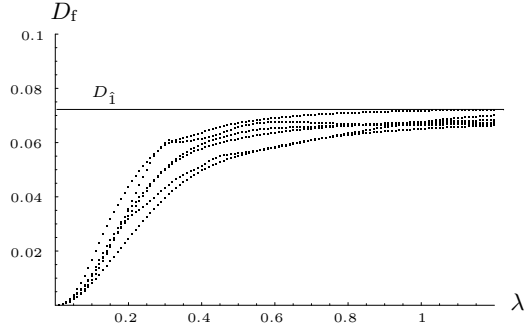,width=7cm}
\caption{Distance $D_\text{f}$ between the expected equilibrium state $\hat{\rho}_{\text{eq}}$ and the final state $\hat{\rho}_{\text{f}}$ in dependence of interaction strength $\lambda$ for six different random interaction in a $N=4$ spin chain.}
\label{figure2}
\end{figure}

As a locally defined thermal equilibrium state for such a system one may expect the (separable) product state:
\begin{equation}
  \label{eq:10}
  \hat{\rho}_{\text{eq}}=\underset{\mu}{\bigotimes}\; \hat{\rho}_{\text{stat}}(\mu).
\end{equation}
To measure the distance between the actual final state $\hat{\rho}_{\text{f}}$ and this expected state $\hat{\rho}_{\text{eq}}$ we use
\begin{equation}
  \label{eq:11}
  D_\text{f}=\Tr\{(\hat{\rho}_{\text{eq}}-\hat{\rho}_{\text{f}})^2\}.
\end{equation}
In FIG.\ \ref{figure2} we vary $\lambda=0.01...1.3$ for different random interactions and evaluate the corresponding distance to the equilibrium state $\hat{\rho}_{\text{eq}}$ in a $N=4$ spin chain.
For decreasing $\lambda$ we find that the resulting $\hat{\rho}_{\text{f}}$ is indeed coming closer to the expected stationary state. 
For very strong interactions we find that the final state is very close to the totally mixed state ($D_{\hat{1}}=\Tr\{(\hat{\rho}_{\text{eq}}-\frac{1}{n^N}\hat{1})^2\}=0.073$).
Qualitatively all the random interactions show the same behavior.

%
%

%
%

However, if we considered instead of the random interaction the case $C_{\text{NR}}=C_{\text{F}}=1$ (F\"orster and non-resonant coupling) we would not get any dependence on the internal interaction strength $\lambda$: 
Irrespective of $\lambda$ all spins of the system have exactly the same temperature as the heat bath. 

In conclusion we always get thermodynamical behavior for the special case of F\"orster and non-resonant coupling. Generally (random interaction), thermodynamical behavior can be expected in the weak coupling limit. 

%
%
\section{Two Baths}

Now we go on to study non-equilibrium properties.
For this purpose we consider again the chain of $N=4$ spins, but now coupled at both ends to separate baths of different temperature but the same $\lambda^{(\text{B})}=\lambda^{(\text{B1})}=\lambda^{(\text{B2})}$. 
For the higher temperature bath B1 we set $W_{0\rightarrow 1}=0.4 \lambda^{(\text{B})}$, $W_{1\rightarrow0}=0.6 \lambda^{(\text{B})}$, which corresponds to a bath temperature in our definition of $T(\text{B1})=0.4$. This bath is coupled to the first spin ($\mu=1$) in the chain.
The lower temperature bath B2 ($T(\text{B2})=0.2$ with $W_{0\rightarrow1}=0.2 \lambda^{(\text{B})}$, $W_{1\rightarrow0}=0.8 \lambda^{(\text{B})}$) is coupled to the last spin ($\mu=4$) in the chain.
Again we assume a weak coupling, $\lambda^{(\text{B})}=0.01$, of the spin system to both baths and vary the internal coupling strength of the system $\lambda$.

\begin{figure}
\psfrag{0.15}{\tiny 0.15}
\psfrag{0.2}{\tiny 0.2}
\psfrag{0.25}{\tiny 0.25}
\psfrag{0.3}{\tiny 0.3}
\psfrag{0.35}{\tiny 0.35}
\psfrag{0.4}{\tiny 0.4}
\psfrag{0.45}{}
\psfrag{1}{\tiny 1}
\psfrag{2}{\tiny 2}
\psfrag{3}{\tiny 3}
\psfrag{4}{\tiny 4}
\psfrag{TB1}{$\scriptscriptstyle T$\tiny(B1)}
\psfrag{TB2}{$\scriptscriptstyle T$\tiny(B2)}
\psfrag{T1}{\hspace{-0.3cm}$T$}
\psfrag{N}{}
\psfrag{N1}{$\mu$}
\psfrag{Bath1}{\tiny hot bath }
\psfrag{Bath2}{\hspace{-0.5cm} \tiny cold bath}
\epsfig{file=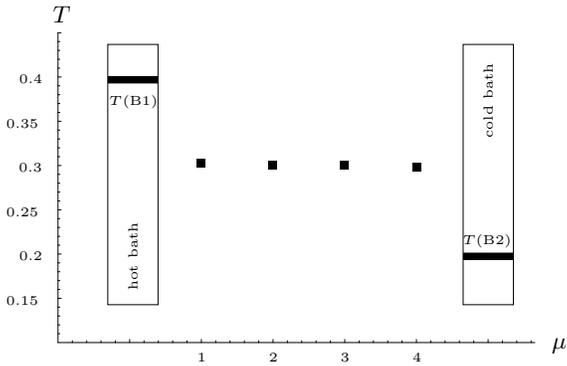,width=8cm}
\caption{Temperature profile: spin chain ($N=4$) with only internal next neighbor F\"orster coupling ($C_{\text{NR}}=0$, $C_{\text{F}}=1$) coupled to heat baths at both ends (lines in the bath boxes mark the appropriate temperature of the bath $T$(B1)=0.4, $T$(B2)=0.2); internal and external coupling strength $\lambda=\lambda^{(\text{B})}=0.01$.}
\label{figure3}
\end{figure}

Here we investigate all three different types of internal couplings between the spins to test the dependence of the interaction. 
Firstly we consider the coupling due to an energy transfer only (F\"orster interaction $\mathcal{H}^{\text{(F)}}$, $C_{\text{F}}=1$, $C_{\text{NR}}=0$). 
For a strong coupling within the chain, $\lambda=1$, all spins get the same averaged temperature between those of the heat baths, independently of the initial state. 
In case of a weak internal interaction strength, $\lambda=0.01$, the two borderline spins are drawn to the temperatures of their bath ($T(1)=0.302$ and $T(4)=0.298$, respectively cf.\ FIG.\ \ref{figure3}). The two spins in the middle are exactly on the same average temperature $T(2)=T(3)=0.3$ independently of the interaction strength (see FIG.\ \ref{figure3}). 
Thus, in the middle of the system the conductivity diverges and Fourier's Law is not fulfilled. 
This is a well known behavior for systems with only energy transfer coupling \cite{Lepri1998,Saito2000}. 

%
%

\begin{figure}
\psfrag{0.15}{\tiny 0.15}
\psfrag{0.2}{\tiny 0.2}
\psfrag{0.25}{\tiny 0.25}
\psfrag{0.3}{\tiny 0.3}
\psfrag{0.35}{\tiny 0.35}
\psfrag{0.4}{\tiny 0.4}
\psfrag{0.45}{}
\psfrag{T1}{\hspace{-0.3cm}$T$}
\psfrag{N1}{$\mu$}
\psfrag{TB1}{$\scriptscriptstyle T$\tiny(B1)}
\psfrag{TB2}{$\scriptscriptstyle T$\tiny(B2)}
\psfrag{Bath1}{\tiny hot bath }
\psfrag{Bath2}{\hspace{-0.5cm}\tiny cold bath}
\psfrag{A1}{$\scriptstyle \blacktriangle\; N=6$}
\psfrag{A2}{$\scriptscriptstyle \blacksquare\scriptstyle\; N=4$}
\epsfig{file=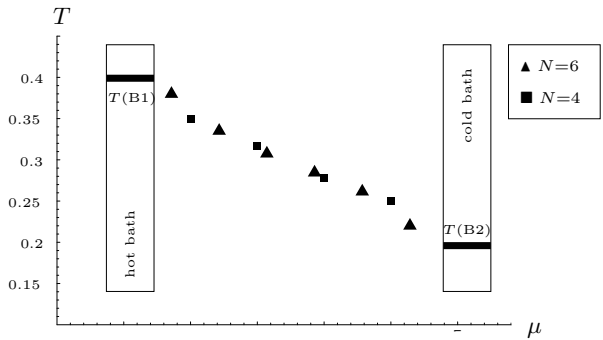,width=8cm}
\caption{Temperature profile: open system with $N=4$ and $N=6$ spins coupled with internal next neighbor F\"orster and non resonant coupling ($C_{\text{NR}}=1$, $C_{\text{F}}=1$); internal and external coupling strength $\lambda=\lambda^{(\text{B})}=0.01$; bath temperatures $T$(B1)=0.4, $T$(B2)=0.2.}
\label{figure4}
\end{figure}

Now we couple the subsystems additionally with a non resonant interaction, which in itself does not give rise to any energy exchange ($\mathcal{H}^{\text{(NR)}}$, $C_{\text{NR}}=1$, $\mathcal{H}^{\text{(F)}}$, $C_{\text{F}}=1$). In this special case we find a non vanishing temperature profile independent of the initial state. In FIG.\ \ref{figure4} we show the temperature profile for a $N=4$ ($N=6$) spin system for $\lambda=\lambda^{(\text{B})}=0.01$.
For this special coupling type, we observe the same profile even for larger $\lambda=\lambda^{(\text{B})}<=10$.
In case of stronger internal coupling ($\lambda^{(\text{B})}=0.01$) we find again a linear profile but with a smaller gradient. 
Summarizing in all cases of F\"orster coupling together with a non-resonant coupling a non vanishing temperature gradient is reached, independently of the internal coupling strength.

%
%

\begin{figure}
\psfrag{0.15}{\tiny 0.15}
\psfrag{0.2}{\tiny 0.2}
\psfrag{0.25}{\tiny 0.25}
\psfrag{0.3}{\tiny 0.3}
\psfrag{0.35}{\tiny 0.35}
\psfrag{0.4}{\tiny 0.4}
\psfrag{0.45}{\tiny 0.45}
\psfrag{0.5}{\tiny 0.5}
\psfrag{1}{\tiny 1}
\psfrag{2}{\tiny 2}
\psfrag{3}{\tiny 3}
\psfrag{4}{\tiny 4}
\psfrag{T1}{\hspace{-0.5cm}$T$}
\psfrag{N1}{$\mu$}
\psfrag{TB1}{$\scriptscriptstyle T$\tiny(B1)}
\psfrag{TB2}{$\scriptscriptstyle T$\tiny(B2)}
\psfrag{Bath1}{\tiny hot bath }
\psfrag{Bath2}{\hspace{-0.5cm}\tiny cold bath}
\psfrag{l2}{$\scriptstyle \blacklozenge\; \lambda=0.1$}
\psfrag{l1}{$\scriptscriptstyle \blacksquare\scriptstyle\; \lambda=0.01$}
\psfrag{l3}{$\scriptstyle \blacktriangle\; \lambda=10$} 
\psfrag{l4}{$\scriptstyle \bigstar\; \lambda=100$} 
\epsfig{file=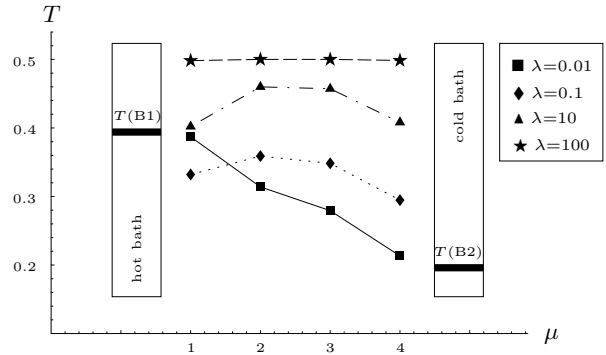,width=8cm}
\caption{Temperature profile: open system with internal random next neighbor coupling for increasing internal coupling strength $\lambda$ ($\lambda^{(\text{B})}=0.01$); bath temperatures $T$(B1)=0.4, $T$(B2)=0.2.}
\label{figure5}
\end{figure}

Finally, the internal interactions of the system are taken to be random next neighbor couplings.
In FIG.\ \ref{figure5} temperature profiles for different internal coupling strength $\lambda$ are shown. 
For decreasing $\lambda$ the profiles approach a linear dependence with finite gradient.
 
%
%

Increasing $\lambda$ leads to a deviation from this linear profile:
In case of very strong internal couplings the system is strongly entangled and therefore we find the spins locally in a totally mixed state.
This behavior is similar for any random interaction.  

\begin{figure}
\psfrag{0.1}{\tiny 0.1}
\psfrag{0.08}{\tiny 0.08}
\psfrag{0.06}{\tiny 0.06}
\psfrag{0.04}{\tiny 0.04}
\psfrag{0.02}{\tiny 0.02}
\psfrag{0.2}{\tiny 0.2}
\psfrag{0.8}{\tiny 0.8}
\psfrag{0.6}{\tiny 0.6}
\psfrag{0.4}{\tiny 0.4}
\psfrag{1}{\tiny 1}
\psfrag{D}{$D_\text{f}$}
\psfrag{D1}{$\scriptstyle D_{\hat{1}}$}
\psfrag{l}{$\lambda$}
\epsfig{file=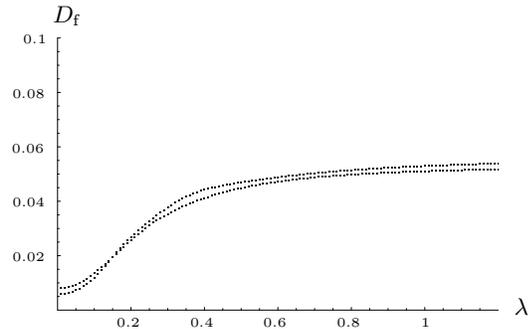,width=7cm}
\caption{Distance $D_\text{f}$ between the expected equilibrium state $\hat{\rho}_{\text{eq}}$ and the final state
$\hat{\rho}_{\text{f}}$ as a function of the interaction strength $\lambda$ for two different random interaction in a $N=4$ spin system.}
\label{figure6}
\end{figure}

The distance between the expected stationary state and the actual final state of the system in dependence of internal coupling strength $\lambda$ can be found in FIG.\ \ref{figure6}($N=4$). Again, the distance increases for a stronger internal coupling. But even for very small interactions we do not get exactly the expected stationary state: Any specific random coupling leads to a somewhat different but approximate linear temperature profile.

%
%

%
%
\section{Flux}

In this section we examine the properties of the energy flux through the
considered spin chain. We consider the change of energy of the single spin $\mu$ using the
Liouville equation (\ref{eq:5}).
\begin{equation}
\langle 1_{\mu} \vert \textrm{Tr}_{\nu}\left\{ \dod{\hat{\rho}}{t} \right \} \vert 1_{\mu} \rangle
= \langle 1_{\mu} \vert \textrm{Tr}_{\nu}\left\{
- \im [\mathcal{H},\hat{\rho}]+\mathcal{L}^{(\text{B})}\hat{\rho}\right \} \vert 1_{\mu} \rangle.
\end{equation}
Here, $\vert 1_{\mu} \rangle$ is the exited state of spin $\mu$ and $\textrm{Tr}_{\nu}$ means
the trace over all other spins $\nu$ ($\nu\not=\mu$).
For a spin in the middle of the chain ($\mu\not=1$ and $\mu\not=N$) the equation reads
\begin{eqnarray}
\dod{\langle 1_{\mu} \vert \hat{\rho} \vert 1_{\mu} \rangle}{t} = 2 \lambda C_{\text{F}}
& \left( \textrm{Im} \langle 1_{\mu-1} 0_{\mu} \vert \hat{\rho} \vert 0_{\mu-1} 1_{\mu} \rangle \right.
\nonumber \\
- & \left.\textrm{Im} \langle 1_{\mu} 0_{\mu+1} \vert \hat{\rho} \vert 0_{\mu} 1_{\mu+1} \rangle \right).
\end{eqnarray}
where trace over all the other spins is assumed. We interpret the quantity
\begin{equation}
J_{\mu, \mu+1} =
2 \lambda C_{\text{F}} \textrm{Im} \langle 1_{\mu} 0_{\mu+1} \vert \hat{\rho} \vert 0_{\mu} 1_{\mu+1} \rangle
\end{equation}
as the flux from spin $\mu$ to spin $\mu+1$ and the above equation shows that for the stationary state,
the flux is the same for every $\mu$, $J =J_{\mu, \mu+1}$, as expected.

\begin{figure}
\psfrag{0.01}{\tiny \hspace{3mm} 1}
\psfrag{0.02}{\tiny \hspace{3mm} 2}
\psfrag{0.03}{\tiny \hspace{3mm} 3}
\psfrag{0.04}{\tiny \hspace{3mm} 4}
\psfrag{0.05}{\tiny \hspace{3mm} 5}
\psfrag{0.1}{\tiny 0.1}
\psfrag{0.15}{\tiny 0.15}
\psfrag{0.2}{\tiny 0.2}
\psfrag{0.25}{\tiny 0.25}
\psfrag{0.3}{\tiny 0.3}
\psfrag{0.5}{\tiny 0.05}
\psfrag{D}{$\scriptstyle \nabla T$}
\psfrag{Fluss}{\hspace{-2mm}$J/10^{-3}$}
\epsfig{file=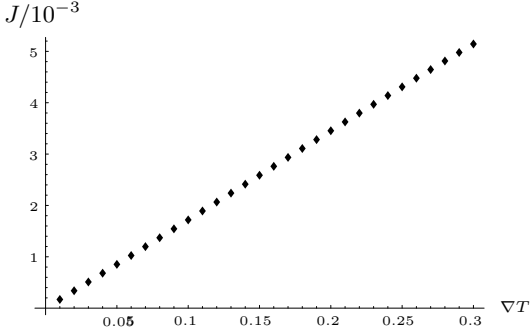,width=7cm}
\caption{Fourier's Law: flux $J$ over temperature gradient $\nabla T$ for a $N=4$ spin chain with F\"orster and non-resonant internal coupling ($C_{\text{F}}=1$, $C_{\text{NR}}=1$, $\lambda=0.1$).}
\label{figure7}
\end{figure}

\begin{figure}
\psfrag{n6}{\tiny $\frac{1}{6}$}
\psfrag{n5}{\tiny $\frac{1}{5}$}
\psfrag{n4}{\tiny $\frac{1}{4}$}
\psfrag{n3}{\tiny $\frac{1}{3}$}
\psfrag{0.001}{\tiny \hspace{4.5mm} 1}
\psfrag{0.002}{\tiny \hspace{4.5mm} 2}
\psfrag{0.003}{\tiny \hspace{4.5mm} 3}
\psfrag{0.004}{\tiny \hspace{4.5mm} 4}
\psfrag{0.005}{\tiny \hspace{4.5mm} 5}
\psfrag{0.006}{}
\psfrag{N}{$\frac{1}{N}$}
\psfrag{j}{\hspace{-5mm}$J/10^{-3}$}
\epsfig{file=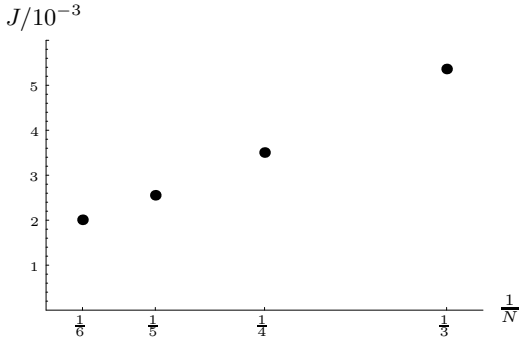,width=7cm}
\caption{The dependence of the flux $J$ on the inverse chain length $N^{-1}$ for $C_{\text{F}}=1$, $C_{\text{NR}}=1$, $\lambda=0.1$
and $N=3,4,5,6$.}
\label{figure8}
\end{figure}

We now test two important properties of the flux, its dependence on the temperature gradient and on the
length of the chain. Again we consider now a system with a F\"orster and a non-resonant coupling. In FIG.\ \ref{figure7} we show the flux $J$ in dependence of the temperature gradient $\nabla T$, the system fulfills Fourier's law
\begin{equation}
  \label{eq:12}
  \text{const}=J= \kappa\nabla T.
\end{equation}
Only if $\nabla T$ is constant we can deduce the conductance $\kappa$ to be a constant material property over the homogeneous ``wire''.
For the same $T$(B1), $T$(B2) the flux $J$ is found to decrease with the chain length $N$: $J\propto\frac{1}{N}$ (see FIG.\ \ref{figure8}), which underlines the fact that the conductance $\kappa$ is a bulk effect, not contact property.
 
%
%

%
%

%
%
\section{Conclusion}

We have studied a one-dimensional quantum system, a spin chain, coupled to heat baths within a Lindblad formalism. 
For given local bath couplings, different interactions between the subsystems in the chain have been introduced, an energy transfer coupling (F\"orster coupling), a non-resonant coupling and an random ``unbiased'' interaction.
The chain properties have been characterized by the profile of the local mean energy and the energy-flux, $J$, respectively. 

In case of coupling to only one bath the system has been found to behave thermodynamical (i.e.\ flat temperature profile, controlled by bath)  if we use the F\"orster coupling together with a non-resonant coupling. For random couplings we find a thermodynamical behavior for small interactions within the system only. 

The scenario for heat conduction, i.e.\ the spin chain coupled between a reservoir with high temperature and a reservoir with  low temperature, shows different behavior in dependence of the type of the internal interactions. For energy transfer coupling only the heat conductivity diverges. With an additional non-resonant interaction we find a non vanishing temperature gradient. For longer chains the system shows the expected scaling properties. In case of small random interactions we always get a  non vanishing temperature gradient, but for increasing interaction strength local thermodynamical behavior is violated.

Even a class of small systems with just four spins may thus show local thermodynamical behavior under quantum mechanical modeling. 
For weakly coupled systems the emergence of a non vanishing temperature gradient appears to be generic if and only if the internal interaction contains some non-resonant part.
For systems with strong internal interaction, however, the local mean energy profile typically shows significant deviations from linear behavior. 
In this case the definition of a local temperature becomes ambiguous.
However, we hope to overcome this weak coupling limitation by introducing a coarse graining, i.e.\ by grouping a number of subsystems together such that these larger blocks exhibit effective weak coupling. 
Then a local energy profile on this effective reduced scale could be defined for which we expect to recover thermodynamical behavior.
This would give an answer to the question on what local scale a temperature or local conductance could reasonably be defined.

There are special interaction models (F\"orster coupling together with non-resonant coupling), for which the profile does not depend on the interaction strength, a thermodynamical behavior can always be observed. 
For energy transfer only (F\"orster coupling) thermodynamical behavior can never be found.
In this way it is possible to characterize the behavior of a large class of small systems according to their internal coupling.

Other open problems relate to the dependence of the heat conductivity $\kappa$ on temperature. In the classical phonon theory of heat conduction in insulators there are some expectations on this dependence based on experimental results \cite{Ziman2001}. We hope to find a corresponding behavior in a fully quantum mechanical treatment. 

\begin{acknowledgement}
We thank P.~Borowski, H.~Schmidt, M.~Stollsteimer, and F.~Tonner for fruitful discussions. 
\end{acknowledgement}

%
%

\end{document}